\documentclass[aps,prl,showpacs,amsmath,amssymb,twocolumn,groupedaddress]{revtex4-1}
\usepackage{graphicx}
\usepackage[colorlinks,linkcolor=blue,citecolor=blue]{hyperref}
\begin{document}


\title{Conceptual Proposal: Frequency Offset Modulation for High-Efficiency Communications}


\author{Xihua Zou}
\email[]{zouxihua@swjtu.edu.cn}
\author{Wei Pan}
\email[]{wpan@swjtu.edu.cn}
\author{Ge Yu}
\author{Bin Luo}
\author{Lianshan Yan}

\affiliation{School of Information Science and Technology, Southwest Jiaotong Unviersity, Chengdu 611756, China}


\date{November, 2016}

\begin{abstract}
Frequency offset modulation (FOM) is proposed as a new concept to provide both high energy efficiency and high spectral efficiency for communications. In the FOM system, an array of transmitters (TXs) is deployed and only one TX is activated for data transmission at any signaling time instance. The TX index distinguished by a very slight frequency offset among the entire occupied bandwidth is exploited to implicitly convey a bit unit without any power or signal radiation, saving the power and spectral resources. Moreover, the FOM is characterized by removing the stringent requirements on distinguishable spatial channels and perfect \textit{priori} channel knowledge, while retaining the advantages of no inter-channel interference and no need of inter-antenna synchronization. In addition, a hybrid solution integrating the FOM and the spatial modulation is discussed to further improve the energy efficiency and spectral efficiency. Consequently, the FOM will be an enabling and green solution to support ever-increasing high-capacity data traffic in a variety of interdisciplinary fields.
\end{abstract}


\maketitle

Exponential increase of data traffic unprecedentedly drives both optical fiber and wireless communications towards huge capacity, massive access and low energy consumption. In wireless communication (e.g., future 5G and beyond), numerous innovative techniques are proposed to improve the capacity, energy efficiency and spectral efficiency, such as multiple-input multiple-output (MIMO) architecture, massive antennas, beam forming, spatial modulation, millimeter-wave technique and radio-over-fiber technology \cite{Soldani2015Horizon,Ren2013Wireless,Roh2014Millimeter,Rusek2013Scaling,Larsson2014Massive,Cao2016Advanced,Bonjour2016Ultra}. For optical fiber communication, coherent techniques supporting orthogonal frequency division multiplexing (OFDM), high-order quadrature amplitude modulation (QAM) and multiple-dimension multiplexing formats have so far been demonstrated \cite{VanWiggeren1998Optical,Essiambre2008Capacity,Li2009Recent,Chen2016Use}.

Multiple antennas, multiple transmitters (TXs) or multiple paths have been configured in MIMO and massive-antenna systems to provide high data rate, high energy efficiency and high spectral efficiency \cite{Paulraj1994Increasing,Foschini1996Layered,Raleigh1996Spatio,Henty2004Multipath,Mesleh2008Spatial}. Among these systems, in particular, the spatial modulation exploits the spatial position of TXs to implicitly convey some bit units without signal radiation. In addition to high energy efficiency, the spatial modulation is also characterized by no inter-channel interference (ICI) and no need of inter-antenna synchronization (IAS) \cite{Mesleh2008Spatial,Jeganathan2008Spatial,DiRenzo2011Spatial,DiRenzo2014Spatial,Stavridis2012energy,Stavridis2013Energy}, with respective to conventional MIMO. To identify the spatial position of the active TX in the spatial modulation, however, two stringent requirements have to be fulfilled, sufficiently distinguishable spatial channels and perfect \textit{priori} channel knowledge \cite{DiRenzo2011Spatial,DiRenzo2014Spatial}. The first requirement might bring limitations on size reduction of TXs and antenna distribution. The second will make the system vulnerable to rapid channel fading or change, particularly in highly dynamic environments and high-mobility scenarios. 

Here, we propose a new concept defined as frequency offset modulation (FOM) for high-efficiency communication, to overcome the issues of the spatial modulation but retaining the advantages. Unlike the spatial modulation, the FOM utilizes a very slight frequency offset among multiple TXs to distinguish the index of the active TX. Consequently, the stringent limitations on identifying the TX index including distinguishable spatial channels and perfect \textit{priori} channel knowledge are fundamentally removed, which would be significantly beneficial for massive-antenna or high-complexity systems and highly dynamic scenarios. Thanks to low power consumption, high spectral efficiency and easy deployment, the FOM will be an enabling solution capable of supporting unprecedentedly growing high-capacity data traffic in various  interdisciplinary fields. 

\begin{figure}[htbp]
\includegraphics[scale=0.65]{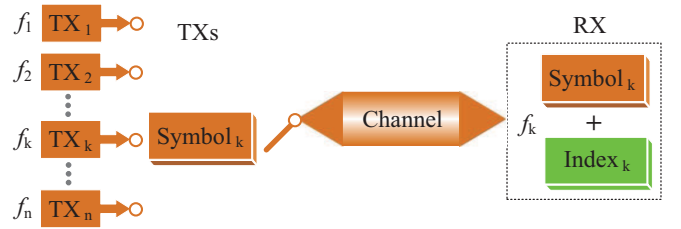}
\caption{Generic architecture of the FOM system.\label{Fig_1}}
\end{figure}

As illustrated in Fig. \ref{Fig_1}, the generic architecture of the FOM system consists of an array of TXs, a channel, and a receiver (RX), which is analog with the architecture of a multiple-input single-output (MISO) or spatial modulation system \cite{Mesleh2008Spatial,DiRenzo2011Spatial}. At each signaling time instance, a single TX is activated to transmit a symbol unit, while all other TXs are kept silent.

\begin{figure}[htbp]
\includegraphics[scale=0.65]{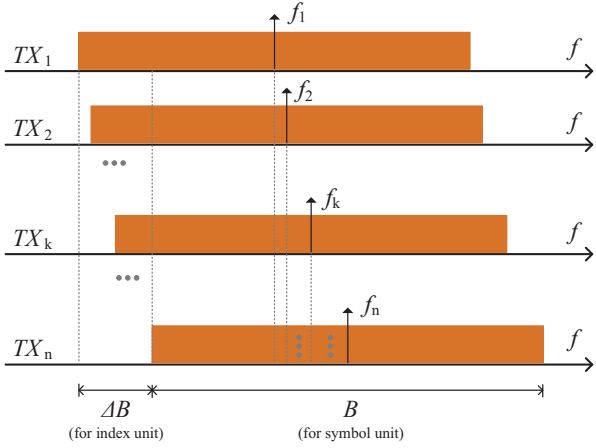}
\caption{Allocation of the central frequencies and the bandwidths for the FOM system.\label{Fig_2}}
\end{figure}

\begin{figure}[htbp]
\includegraphics[scale=0.65]{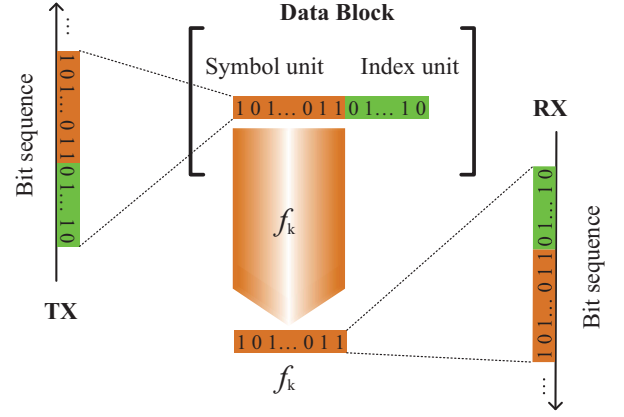}
\caption{Data transmission procedure of the FOM.\label{Fig_3}}
\end{figure}

Here, the novelty of the proposed FOM system lies in the use of a very slight frequency offset to distinguish any active TX in the array. As shown in Fig. \ref{Fig_2}, each TX is allocated to operate at a specific central frequency $f_k$, where $1\leq k\leq n$ and $n$ is the total number of TXs. The large bandwidths occupied by all the TXs are mostly overlapping. There is a very slight frequency offset ($\Delta f$) between two neighboring TXs, such that each TX has a specific frequency offset serving as its fingerprint. Mathematically, the relationships can be expressed as
\begin{align}
\Delta f&=f_2-f_1=f_k-f_{k-1}=f_n-f_{n-1},\label{eq_1}\\
\Delta f&\ll f_k,\label{eq_2}\\
\Delta B&=(n-1)\Delta f,\label{eq_3}\\
\Delta B&\ll B.\label{eq_4}
\end{align}
$B$ and $\Delta B$ are the bandwidths allocated for transmitting symbol units and for accommodating the frequency offset of the TXs, as shown in Figs. \ref{Fig_2} and \ref{Fig_3}. 

As illustrated in Fig. \ref{Fig_3}, a bit sequence is divided into a number of data blocks. Each data block contains two data carrying units, the symbol unit and the index unit. The symbol unit is encoded or mapped by following advanced symbol constellation diagram (e.g., QAM) and then emitted by the active $TX_k$ operating at the central frequency of $f_k$, while the index unit is mapped by the allocated frequency-shift constellation diagram of the active TX ($TX_k$). The symbol unit emitted from $TX_k$ explicitly travels through an optical or wireless channel. At the RX, the index of the active TX can be differentiated through the frequency offset, which is designed to implicitly convey the index unit without radiating any optical or electrical signal. A full data block is then recovered by combining the symbol unit and index unit, as shown in Figs.\ref{Fig_1} and \ref{Fig_3}.  Thus, the FOM is characterized to be energy efficient due to the power-free data transmission of the index unit.

It is also noted that the large bandwidths allocated for all TXs are mostly overlapping, despite a very slight frequency offset among the central frequencies. Thanks to the bandwidth overlapping, the FOM is capable of saving bandwidth to retain a high spectral efficiency very close to that of the spatial modulation.

The energy efficiency and spectral efficiency of the FOM will be characterized in detail. Given a fixed energy consumption for conveying one bit, an energy efficiency ratio ($\Gamma_{EE}$) between the consumed energies in the conventional QAM and in the FOM systems for transmitting ($\log_2m+\log_2n$)  bits can be derived as
\begin{equation}\label{eq_5}
\Gamma_{EE}=\frac{\log_2m}{\log_2m+\log_2n}=\frac{1}{1+\log_mn},
\end{equation}
where $m$ denotes the cardinality of the symbol-constellation diagram (e.g.,$m=8$ for 8-QAM). It can be seen from Eq. (\ref{eq_5}) that the energy consumption is reduced to be $1/(1+\log_mn)$ in the FOM, compared with a QAM system.

In principle, the energy efficiency of the FOM is comparable to that of the spatial modulation. Consequently, according to the results reported in Refs. \cite{Stavridis2012energy,Stavridis2013Energy}, the FOM enables an energy saving by $67\%$ or even up to $90\%$, compared with conventional MIMO systems. Moreover, the energy efficiency of the FOM or the spatial modulation would be further optimized with the increase of TXs or paths.

From Fig. \ref{Fig_2}, the spectral efficiency ($E_0$) corresponding to the symbol unit can be expressed as
\begin{equation}\label{eq_6}
E_0=\frac{R\times\log_2m}{B},
\end{equation}
where $R$ is the symbol rate. In the FOM system, the spectral efficiency ($E_s$)  for the entire data block (i.e., symbol unit and index unit) is derived as
\begin{equation}\label{eq_7}
E_s=\frac{R\times(\log_2m+\log_2n)}{B+\Delta B}.
\end{equation}
In general,$E_s\geq E_0$ should be fulfilled for providing an improved spectral efficiency, together with high energy efficiency. A spectral efficiency ratio between (\ref{eq_6}) and (\ref{eq_7}) can be defined as
\begin{equation}\label{eq_8}
\Gamma_{SE}=\frac{E_s}{E_0}=\frac{1+\log_mn}{1+\eta},
\end{equation}
where $\eta=\Delta B/B$. A high spectral efficiency can be retained or enhanced as long as $\Gamma_{SE}\geq 1$. Accordingly, an more explicit inequality shown in Eq. (\ref{eq_9}) should be satisfied to ensure a spectral efficiency no less than that of the symbol unit (i.e., the original QAM system).
\begin{equation}\label{eq_9}
n\geq m^\eta.
\end{equation}

\begin{figure}[htbp]
\includegraphics[scale=0.5]{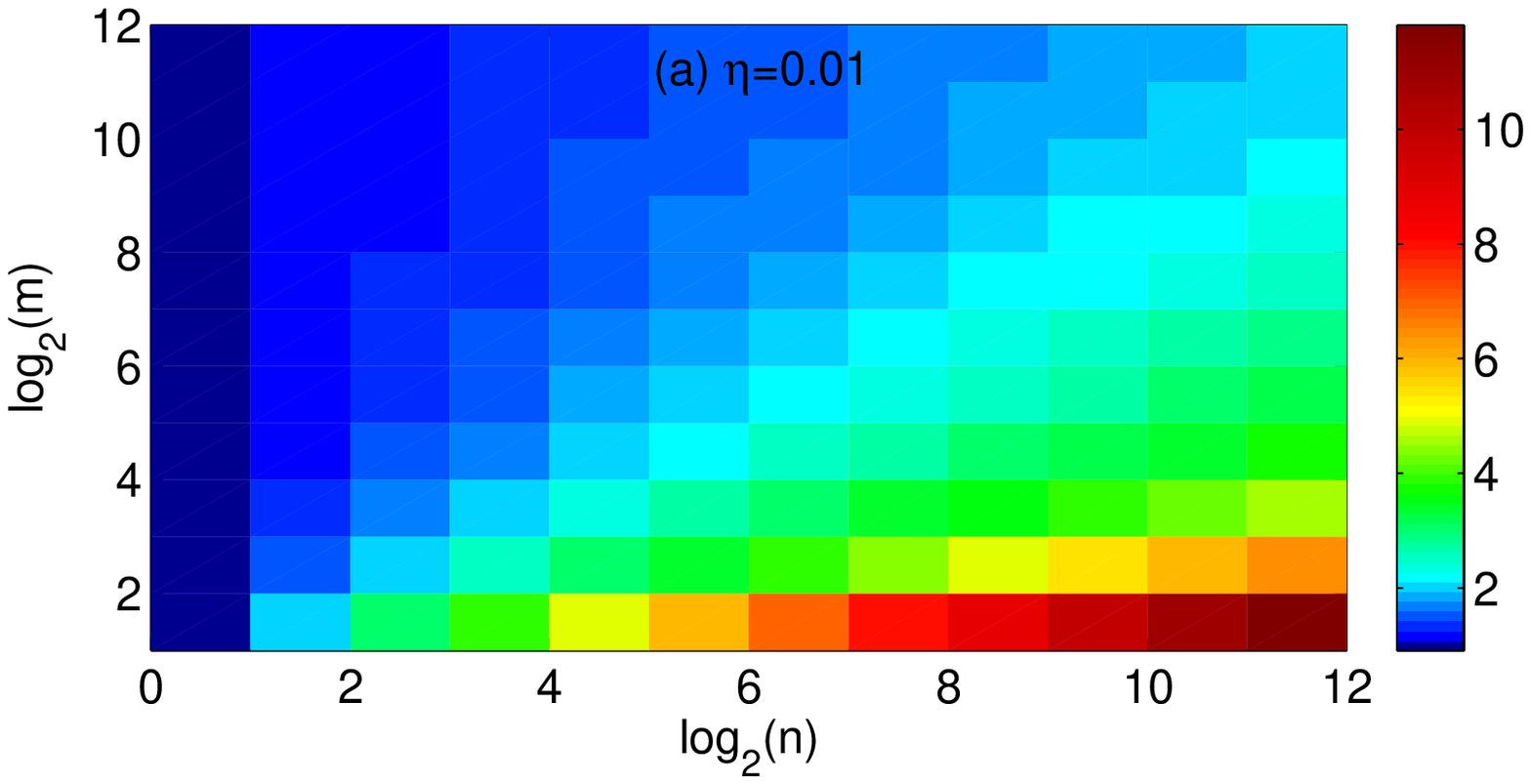}
\includegraphics[scale=0.5]{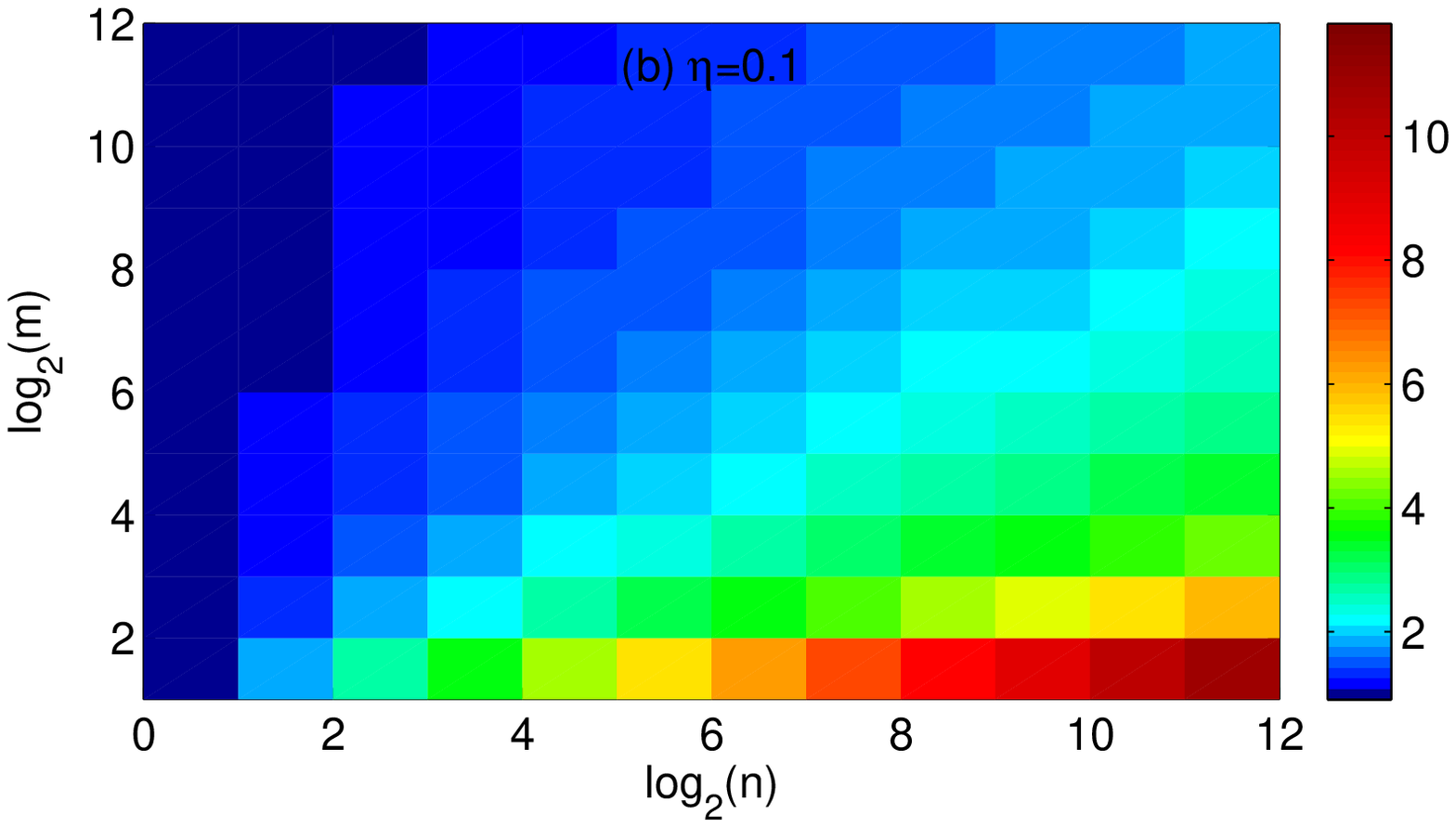}
\caption{Estimated spectral efficiency ratio with a fixed $\Delta B$:(a)$\eta=0.01$ and (b)$\eta=0.1$\label{Fig_4}}
\end{figure}

Quantitative analysis on the spectral efficiency is demonstrated in two cases. For a fixed $\Delta B$ or $\eta$, the spectral efficiency ratio can be analyzed by adjusting the number ($n$) of TXs and the cardinality ($m$) of the symbol constellation. As shown in Fig. \ref{Fig_4}(a), with a fixed $\eta=0.01$, only in a small area (dark blue on the left) $\Gamma_{SE}$ is less than or close to 1 when $n=1$. Most spectral efficiency ratios are estimated to be greater than 1, such as $\Gamma_{SE}=1.86$ for $m=256$ and $n=128$. The spectral efficiency can be further improved by increasing the number of TXs. But a more complex constellation to the symbol unit will relax the increase of the spectral efficiency, such as 1024-QAM. Further, if $\Delta B$ is increased to have $\eta=0.1$, an improved spectral efficiency is still achieved in most areas of the color gradient chart shown in Fig. \ref{Fig_4}(b), despite a small dark-blue area on the upper left for $\Gamma_{SE}$ less than or close to 1.

\begin{figure}[htbp]
\includegraphics[scale=0.56]{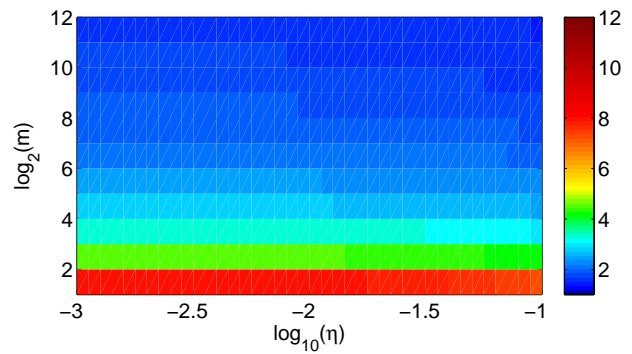}
\caption{Estimated spectral efficiency ratio for a fixed $n=128$ \label{Fig_5}}
\end{figure}

For a fixed $n$, the spectral efficiency ratio can be analyzed by adjusting the bandwidth ($\Delta B$) allocated for the TX index of the FOM  and the cardinality ($m$) of the symbol constellation. As shown in Fig. \ref{Fig_5}, all values are estimated to be larger than 1.44 and even up to 8 for $0.001\leq \eta\leq 0.1$, when $n=128$. 


\begin{figure}[htbp]
\includegraphics[scale=0.6]{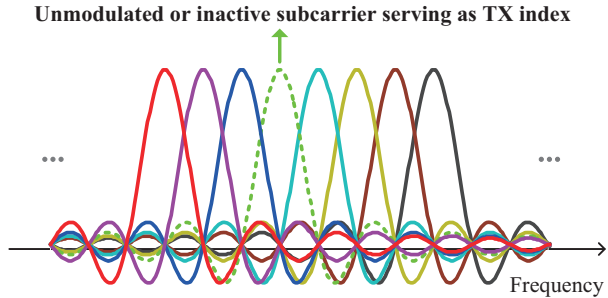}
\caption{Equivalent index allocation for the FOM in a parallel combinatory or subcarrier-index modulation OFDM system.  \label{Fig_6}}
\end{figure}

 Here, two equivalent solutions are presented as proof-of-concept demonstrations in the following. To implement the FOM, available ultralow-expansion cavities with linewidth at kilohertz level \cite{Ludlow2007Compact,Doringshoff2010High,Hirata2014Sub} are able to provide high-resolution frequency discrimination, but at the cost of increasing complexity of the RX. Furthermore, the principle of FOM can be equivalently demonstrated in a parallel combinatory OFDM \cite{Sasaki1995Multiple,Frenger1999Parallel,Kitamoto2005Parallel,Yafei2007novel} or subcarrier-index modulation OFDM \cite{Abu-Alhiga2009Subcarrier,Tsonev2011Enhanced,Basar2013Orthogonal,Ishikawa2016Subcarrier,Ma2016Subcarrier}, when a symbol unit contains one symbol. As shown in Fig. \ref{Fig_6}, the total bandwidth occupied by the multiple subcarriers of the OFDM is equal to that allocated for the FOM (i.e., $B+\Delta B$) and one subcarrier of the OFDM is kept being unmodulated or inactive to equivalently function as the index of an active TX. Here, the narrow bandwidth $\Delta B$ allocated for the TX index unit is not a fixed range any more, but a flexible one among the entire bandwidth ($B+\Delta B$). In a typical OFDM system (e.g, a 20-MHz bandwidth and a 15-kHz spacing \cite{Dahlman20144G}), the number of the TXs is specified to be identical with the number of the subcarriers, such that the preconditions [Eqs. (\ref{eq_2}) and (\ref{eq_4})] of the FOM are usually fulfilled. At each signaling time instance, one subcarrier is kept unmodulated or inactive to implicitly transmit the index unit. Therefore, the subcarrier (i.e., the index of TX) can effectively distinguished by using available signal processing techniques in the parallel combinatory or subcarrier-index modulation OFDM.  It should be noted that a high spectral efficiency can still be retained under the condition that the number of the subcarriers is larger than the cardinality of QAM constellation diagram (i.e., $n\geq m$) which is valid for most OFDM systems.

According to the principle and the analysis above, it is clear that the FOM has all the advantages of the spatial modulation, including the promotion of the energy efficiency and spectral efficiency, the elimination of ICI, and no need of IAS, with respective to conventional MIMO or single-input single-output (SISO) systems. More importantly, the index or fingerprint of the active TX is distinguished through a very slight frequency offset which is long-term stable and insensitive to dynamic channels or environments. As a result, the FOM is capable of greatly mitigating the two stringent requirements facing the spatial modulation. Namely, distinguishable spatial channels and perfect \textit{priori} channel knowledge are no longer needed for identifying the TX index, making the FOM more competitive for highly integrated frontends, massive TXs in FOM and highly dynamic scenarios. On the other hand, 
a new challenge facing the FOM is the implementation of precise frequency offset estimation in a symbol, due to the tradeoff between a limited symbol duration and a required high frequency resolution. Therefore, identifying the index of the active TX is the crucial step to a full implementation of the FOM system. Fortunately, the advances for precise subcarrier frequency identification in the subcarrier-index modulation OFDM \cite{Abu-Alhiga2009Subcarrier,Tsonev2011Enhanced,Basar2013Orthogonal,Ishikawa2016Subcarrier,Ma2016Subcarrier} will greatly facilitate the implementation of the FOM system. At the mean time, ultra-fast symbol-by-symbol processing has been demonstrated as a potential for high data rate and massive access in wireless communication \cite{Bonjour2016Ultra}.

In addition to the fundamental architecture shown in Fig. \ref{Fig_1}, an extended version of the proposed FOM is discussed here. A hybrid system seamlessly integrating the FOM with the spatial modulation can be implemented to further improve the energy efficiency and spectral efficiency. A two-dimension (2D) index including the frequency offset of the FOM and the spatial position of the spatial modulation can be simultaneously formed for each TX. When both the number of the TXs and the number of central frequencies are specified as $n$, the index unit of each data block contains $2\log_2n$ bits which are twice that of an FOM or spatial modulation system. Consequently, the energy efficiency and spectral efficiency can be improved as,
\begin{align}
\Gamma_{EE}&=\frac{1}{1+2\log_mn},\\
\Gamma_{SE}&=\frac{1+2\log_mn}{1+\eta}.
\end{align}

The FOM, as a new concept, is proposed for high-efficiency communications. Each TX is distinguished by a very slight frequency offset and then some bits are implicitly transmitted through the TX index without radiating any signal. The FOM is able to remove stringent requirements including highly distinguishable spatial channels and perfect \textit{priori} channel knowledge while retaining high energy efficiency and high spectral efficiency, with respective to the spatial modulation. Analysis on the energy efficiency and spectral efficiency are demonstrated and then an extended version with 2D TX index is also discussed. Many efforts are urgently needed for theory modeling, frequency-offset identification, optimized detection solution, experiment verification and practical deployment, to pave the way for the applications of the FOM in a variety of interdisciplinary fields.

\begin{acknowledgments}
This work was supported in part by the ``863'' Project (2015AA016903) and the National Natural Science Foundation of China (61378008). X. Zou was supported by the Research Fellowship of the Alexander von Humboldt Foundation, Germany.
\end{acknowledgments}

\bibliography{fom}

\end{document}